\documentclass[DIV=calc,paper=a4,fontsize=11pt,twocolumn]{scrartcl} % KOMA-article class

\pdfoutput=1
\pdfmapfile{+txfonts.map}

\usepackage[english]{babel}
\usepackage[protrusion=true,expansion=true]{microtype}
\usepackage{amsmath,amsfonts,amsthm}
\usepackage[final]{graphicx}
\usepackage{xcolor}
\usepackage[normal,small,hypcap,up,labelfont=bf,textfont=it]{caption}
\usepackage{epstopdf}
\usepackage{subfig}
\usepackage{booktabs}
\usepackage{fix-cm}
\usepackage{amssymb,amsfonts}
\usepackage{dsfont}
\usepackage{bbm}
\usepackage{pstricks}
\usepackage{cite}
\usepackage[utf8]{inputenc}
\usepackage[perpage,symbol*]{footmisc}
\usepackage[varg]{txfonts}
\usepackage{balance}
\usepackage{fancyhdr}
\PassOptionsToPackage{hyphens}{url}\usepackage[pdfencoding=auto,psdextra]{hyperref}
\usepackage{bookmark}
\usepackage{verbatim}

\usepackage{cuted}
\usepackage{widetext}
\usepackage{slashed}
\usepackage{fontenc}
\usepackage{bm}
\usepackage{mathrsfs}

\theoremstyle{definition}

\theoremstyle{plain}

\DeclareCaptionFont{mycolor}{\color[HTML]{000000}}
\usepackage{sectsty}													% Custom sectioning (see below)
\allsectionsfont{%															% Change font of al section commands
\color[HTML]{31ADF3}\usefont{OT1}{phv}{b}{n}%										% bch-b-n: CharterBT-Bold font
}

\sectionfont{%																% Change font of \section command
\color[HTML]{31ADF3}\usefont{OT1}{phv}{b}{n}%										% bch-b-n: CharterBT-Bold font
}

\usepackage{fancyhdr}												% Needed to define custom headers/footers
\usepackage{lettrine}
\newcommand{\initial}[1]{%
\lettrine[lines=3,lhang=0.3,nindent=0em]{
\color[HTML]{31ADF3}
{\textsf{#1}}}{}}

\usepackage{titling}															% For custom titles

\newcommand{\HorRule}{\color[HTML]{31ADF3}%			% Creating a horizontal rule
\rule{\linewidth}{1pt}%
}

\pretitle{\vspace{-30pt} \begin{flushleft} \HorRule
\fontsize{34}{34} \usefont{OT1}{phv}{b}{n} \color[HTML]{31ADF3} \selectfont
}
\title{How Dirac's Seminal Contributions Pave the Way for Comprehending Nature's Deeper Designs}					% Title of your article goes here
\posttitle{\par\end{flushleft}\vskip 0.5em}

\preauthor{\begin{flushleft}\large \lineskip 0.5em \usefont{OT1}{phv}{b}{sl} \color[HTML]{31ADF3}}
\author{Mani L. Bhaumik\\[8pt]}											% Author names go here
\postauthor{\footnotesize \usefont{OT1}{phv}{m}{sl} \color[HTML]{000000}
Department of Physics, University of California, Los Angeles. E-mail: \href{mailto:bhaumik@physics.ucla.edu}{bhaumik@physics.ucla.edu}\\[10pt]		% Institutions of authors
\scriptsize\usefont{OT1}{phv}{m}{n} \color[HTML]{31ADF3}{\textbf{Editors: \emph{Zvi Bern} \& \emph{Danko Georgiev}} }\\[5pt]
\color[HTML]{000000}{Article history: Submitted on October 27, 2019;  Accepted on December 6, 2019; Published on December 17, 2019.}
\par\end{flushleft}\HorRule}

\date{}																				% No date

\begin{document}
\maketitle
\thispagestyle{fancy} 			% Enabling the custom headers/footers for the first page
% The first character should be within \initial{}
\initial{C}\textbf{redible reasons are presented to reveal that many of the lingering century old enigmas, surrounding the behavior of at least an individual quantum particle, can be comprehended in terms of an objectively real specific wave function. This wave function is gleaned from the single particle energy-momentum eigenstate offered by the theory of space filling universal quantum fields that is an inevitable outcome of Dirac's pioneering masterpiece. Examples of these well-known enigmas are wave particle duality, the de Broglie hypothesis, the uncertainty principle, wave function collapse, and predictions of measurement outcomes in terms of probability instead of certainty. Paul Dirac successfully incorporated special theory of relativity into quantum mechanics for the first time. This was accomplished through his ingenious use of matrices that allowed the equations of motion to maintain the necessary first order time derivative feature necessary for positive probability density. The ensuing Dirac equation for the electron led to the recognition of the mystifying quantized spin and magnetic moment as intrinsic properties in contrast to earlier \emph{ad hoc} assumptions. The solution of his relativistic equation for the hydrogen atom produced results in perfect agreement with experimental data available at the time. The most far reaching prediction of the celebrated Dirac equation was the totally unexpected existence of anti-particles, culminating in the eventual development of the quantum field theory of the Standard Model that reveals the deepest secrets of the universe known to date.\\ Quanta 2019; 8: 88--100.}

\begin{figure}[b!]
\rule{245 pt}{0.5 pt}\\[3pt]
\raisebox{-0.2\height}{\includegraphics[width=5mm]{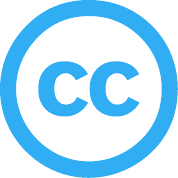}}\raisebox{-0.2\height}{\includegraphics[width=5mm]{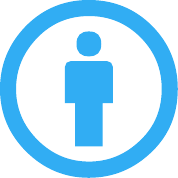}}
\footnotesize{This is an open access article distributed under the terms of the Creative Commons Attribution License \href{http://creativecommons.org/licenses/by/3.0/}{CC-BY-3.0}, which permits unrestricted use, distribution, and reproduction in any medium, provided the original author and source are credited.}
\end{figure}

\section{Introduction}

Just after getting his Ph.D. in 1926, the same year in which physics
started to advance at breakneck speed following the successful formulation
of quantum physics to resolve the perplexities of the atomic domain,
Paul Dirac entered the field of research as a freshly minted scientific
prodigy. Within a mere couple of years, he fashioned an elegant equation,
ever to be known as the iconic Dirac equation of the electron. In
contrast to extensive earlier efforts, his ingenious use of a special
type of non-commutative $4\times4$ matrices allowed the relativistic
quantum mechanical equations of motion to maintain the sought after
hallmark first order time derivative characteristic of the Schr\"{o}dinger
nonrelativistic wave equation. This was a brilliant scheme for successfully
combining the special theory of relativity with quantum mechanics
to explain the behavior of electrons more accurately, avoiding the
various problems encountered in earlier investigations.

First we present an extended overview of the emergence of Dirac's
seminal equation, which had a profound impact for a large variety
of topics including, most importantly, that seeded the eventual development
of the quantum field theory of the Standard Model of particle physics.
We will then focus on how the principles of the quantum field theory
can be utilized for successfully uncovering the century old enigmas
surrounding quantum physics.

In order to fully appreciate Dirac's momentous contribution, it would
be helpful to briefly portray the particularly confounding circumstances
of the period following the epochal unveiling of the Schr\"{o}dinger's
quantum mechanical wave equation. As early as in 1887, while trying
to find a precise standard of length for use in their intended detection
of ether drift, Albert A. Michelson and Edward W. Morley \cite{key-1}
observed that the red hydrogen spectral line is actually a closely
spaced double line. The accuracy of measurement was so impressive
that Michelson received the 1907 physics Nobel Prize for precision
metrology and spectroscopy---and not, as might be commonly assumed,
for the celebrated null result for ether drift that eventually provided
one of the underpinnings for Einstein's formulation of special relativity.

The remarkable effectiveness of incorporating the special theory of
relativity by Arnold Sommerfeld into the old quantum theory, before
Schr\"{o}dinger, was already evident in providing a more accurate
agreement with experimental results of the hydrogen spectrum. Hence
it was natural to seek a relativistic quantum mechanical wave equation.
In fact, in his own attempt to formulate the quantum mechanical wave
equation, Schr\"{o}dinger first constructed a relativistic equation.
He did not, however, publish this formulation since it did not agree
with experimental results.

\section{The Klein--Gordon Equation}

In 1926, a number of authors independently formulated a relativistic
wave equation, although the priority in publication belongs to Oscar
Klein \cite{key-2}. It is known today as the Klein--Gordon equation.

The Klein--Gordon equation is derived using the relativistic mass-energy
equation 
\begin{equation}
E^{2}=m_{0}^{2}c^{4}+p^{2}c^{2}\label{eq:1}
\end{equation}
and drawing on the quantum mechanical energy and momentum operators,
\begin{equation}
E=\imath\hbar\frac{\partial}{\partial t},\qquad p=-\imath\hbar\frac{\partial}{\partial x}\label{eq:2}
\end{equation}
ensuing in 
\begin{equation}
-\hbar^{2}\frac{\partial^{2}}{\partial t^{2}}=m_{0}^{2}c^{4}-\hbar^{2}c^{2}\frac{\partial^{2}}{\partial x^{2}}.\label{eq:3}
\end{equation}
For a wave function $\psi$, dividing equation (\ref{eq:3}) by $\hbar^{2}c^{2}$
and rearranging the terms, the Klein--Gordon equation becomes 
\begin{equation}
\frac{1}{c^{2}}\frac{\partial^{2}}{\partial t^{2}}\psi-\nabla^{2}\psi+\frac{m_{0}^{2}c^{2}}{\hbar^{2}}\psi=0.\label{eq:4}
\end{equation}
The equation (\ref{eq:4}) can be written in a graceful form 
\begin{equation}
\left(\boxdot+\mu^{2}\right)\psi=0
\end{equation}
where $\mu=\frac{m_{0}c}{\hbar}$ and $\boxdot$ is the d'Alembert
operator, 
\begin{equation}
\boxdot=\frac{1}{c^{2}}\frac{\partial^{2}}{\partial t^{2}}-\nabla^{2}.
\end{equation}
Soon the equation became the subject of numerous papers and was considered
by many to be the correct and natural generalization of the Schr\"{o}dinger
equation. However, although it had a mathematical and aesthetic appeal,
before long the limitation of its range of applicability became clear.
To begin with, just like Schr\"{o}dinger's earlier unpublished work,
the equation failed to reproduce Sommerfeld's formula for the hydrogen
spectrum that was in exact agreement with experiments. Possibly one
of the biggest disappointments with the Klein--Gordon equation was
its failure to account for the spin of the electron, as was also true
of the nonrelativistic Schr\"{o}dinger equation. However, it has
been revived for use in the quantum field theory of spin $0$ particles
such as the Higgs boson.

Even before the emergence of the brave new world of contemporary quantum
physics, some experimental observations were very baffling and required
the introduction of electron spin to make sense of them. These are
exemplified by the Stern--Gerlach experiment, the anomalous Zeeman
effect, and Wolfgang Pauli's need for two additional quantum numbers
to complete his Pauli exclusion principle, which provides a natural
explanation for the periodic table of atoms.

In 1922, Otto Stern and Walther Gerlach conducted a rather remarkable
experiment \cite{key-3} to test the Bohr--Sommerfeld model of the
atom of the old quantum theory. A beam of silver atoms having a single
electron in their outer shell was passed through a spatially varying
magnetic field, which deflected them before they struck a detector
screen. Stern and Gerlach found that the electrons were deflected
discretely into only two clusters, a spectacular evidence of space
quantization in an atom, supporting the evolving quantum paradigm
at the time.

Niels Bohr was so startled that he personally wrote to Gerlach
\begin{quote}
I would be grateful if you or Stern could let me know, in a few lines,
whether you interpret your experimental results in this way that the
atoms are oriented only parallel or opposed, but not normal to the
field, as one could provide theoretical reasons for the latter assertion.
\cite{key-4}
\end{quote}
Furthermore, in some experiments with atoms in a magnetic field, spectral
lines were observed to split into four, six, or even more lines and
some triplets showed wider spacing than expected. These deviations
were labeled ``anomalous Zeeman effect'' and were completely incomprehensible
to early investigators.

Meanwhile Pauli devised the Pauli Exclusion Principle in 1925. The
principle states that no two electrons can share the same quantum
state at the same time \cite{key-5}. This means that no two electrons
in a single atom can have the same $n$, $\ell$, $m$, and $s$ numbers.
However, the fourth quantum number was unknown at the time and he
had to introduce it as an arbitrary assumption.

To explain all these conundrums, two graduate students, George Uhlenbeck
and Samuel Goudsmit boldly hypothesized \cite{key-6} that each electron
spins with an angular momentum of one half Planck constant and carries
a magnetic moment of one Bohr magneton. Spin is a quantum property
of electrons without a classical analogue and is a form of angular
momentum. The magnitude of this angular momentum is invariable.

Consequently, spin is now considered to be an intrinsic property of
the electron with quantum number $s$ having values of $s=+\frac{1}{2}$
and $-\frac{1}{2}$. Along with Lorentz, Pauli initially was adamantly
against the concept of rotation of a presumed rigid electron that
could lead to a violation of relativity at the periphery. Goudsmith
was so disappointed that he requested his thesis advisor Paul Ehrenfest
not to submit their paper on spin for publication. Serendipitously,
by then Ehrenfest had already submitted the paper and remarked that
his students were young enough to accept some stupidity.

Eventually, Pauli relented and accepted the notion of the electron
spin, as it gave a persuasive explanation for his two assumed additional
quantum numbers as well as for the anomalous Zeeman Effect. The Stern--Gerlach
experiment was also correctly explained as a consequence of the two
distinct values of the electron spin quantum number. The result of
the Stern--Gerlach experiment has been of abiding interest since no
other experiment demonstrates such a graphic evidence of the quantized
spin.

Pauli also presented a modified Schr\"{o}dinger equation for spin
$\frac{1}{2}$ particles taking into account the interaction of the
particle's spin magnetic moment with an external magnetic field caused
by the orbital motion of the electron \cite{key-7}. This was a somewhat
\emph{ad hoc} modification to the Schr\"{o}dinger equation to explain
the existence of a doublet of additional energy levels attributable
to the presence of spin. The natural occurrence of the electron spin
and its magnetic moment still remained shrouded in a great mystery
and the scientific community continued to devote significant efforts
to find a cogent explanation of the conundrum. The relativistic Klein--Gordon
equation offered no help either.

Perhaps the major concern with the Klein--Gordon equation was that
the probability density given by the equation was not positive-definite.
Apparently Dirac was the first to realize the problem with the probability
interpretation for equations with second-order time derivatives. Also,
because special relativity requires treating time and space on equal
footing, Dirac reasoned that the equation has to be first order not
only in time derivatives, but also in spatial derivatives. Instinctively,
he persevered on finding an equation with these features.

While Dirac was focusing his full attention on the development of
a relativistic quantum equation of motion in first order, to his great
surprise, he did not receive any encouragement from the illustrious
pioneer Niels Bohr. As Dirac recalls
\begin{quote}
I remember once when I was in Copenhagen, that Bohr asked me what
I was working on and I told him that I was trying to get a satisfactory
relativistic theory of the electron, and Bohr said `But Klein and
Gordon have already done that!' That answer first rather disturbed
me. Bohr seemed quite satisfied by Klein's solution, but I was not
because of the negative probabilities that it led to. I just kept
on with it, worrying about getting a theory which would have only
positive probabilities. \cite[p. 690]{key-8}
\end{quote}

\section{The Dirac Equation}

To derive the equation, as usual Dirac started with the relativistic
mass-energy equation (\ref{eq:1}), which using the Pythagoras relation
$p^{2}=p_{x}^{2}+p_{y}^{2}+p_{z}^{2}$ yields
\begin{equation}
E^{2}=m_{0}^{2}c^{4}+\left(p_{x}^{2}+p_{y}^{2}+p_{z}^{2}\right)c^{2}\label{eq:8}
\end{equation}
Getting the equation (\ref{eq:1}) in first order appeared to be a
daunting task to most scientists of the time. Dirac was also very
aware that a number of other physicists were working hard to construct
a satisfactory relativistic quantum theory of the electron. It is
just that time when Dirac's coup came along. As the story goes according
to George Gamow \cite[p. 125]{key-9}, the intuition to achieve the
correct equation flashed in Dirac's mind one evening while he was
staring into the fireplace at St John's College, Cambridge.

With his superb insight, Dirac ingeniously realized that the seemingly
irresolvable conflict between the demands of relativity and his keenly
perceived need for a first order equation in time derivative could
be realized if he could find some matrices $\alpha_{0}$, $\alpha_{1}$,
$\alpha_{2}$, $\alpha_{3}$ that would satisfy the relation, 
\begin{equation}
\alpha_{0}^{2}=\alpha_{1}^{2}=\alpha_{2}^{2}=\alpha_{3}^{2}=1
\end{equation}
and the anti-commutation relations hold
\begin{equation}
\alpha_{i}\alpha_{j}=-\alpha_{j}\alpha_{i},\qquad\left(i\ne j\right).
\end{equation}
He could then construct an equation for energy
\begin{equation}
E=\alpha_{0}m_{0}c^{2}+c\left(p_{x}\alpha_{1}+p_{y}\alpha_{2}+p_{z}\alpha_{3}\right)\label{eq:11}
\end{equation}
By taking the square of the equation (\ref{eq:11}), one could get
the relativistic energy equation (\ref{eq:8}). The matrices $\alpha_{0}$,
$\alpha_{1}$, $\alpha_{2}$, $\alpha_{3}$ are $4\times4$ matrices
as follows
\begin{alignat}{1}
\alpha_{0} & =\left(\begin{array}{cccc}
0 & 0 & 1 & 0\\
0 & 0 & 0 & 1\\
1 & 0 & 0 & 0\\
0 & 1 & 0 & 0
\end{array}\right),\\
\alpha_{1} & =\left(\begin{array}{cccc}
0 & 1 & 0 & 0\\
1 & 0 & 0 & 0\\
0 & 0 & 0 & -1\\
0 & 0 & -1 & 0
\end{array}\right),\\
\alpha_{2} & =\left(\begin{array}{cccc}
0 & -\imath & 0 & 0\\
\imath & 0 & 0 & 0\\
0 & 0 & 0 & \imath\\
0 & 0 & -\imath & 0
\end{array}\right),\\
\alpha_{3} & =\left(\begin{array}{cccc}
1 & 0 & 0 & 0\\
0 & -1 & 0 & 0\\
0 & 0 & -1 & 0\\
0 & 0 & 0 & 1
\end{array}\right).
\end{alignat}
Dirac was unaware of the fact that this type of matrices had been
developed by William K. Clifford in 1877 and even earlier in 1840
by William R. Hamilton to take square root of a second order wave
operator, which was subsequently shown to be subtly relativistically
invariant. To quote Sir Roger Penrose
\begin{quote}
It is perhaps not surprising that Dirac was unaware of Clifford's
discoveries of over half a century earlier, because this work was
not at all well known in 1920s, even to many specialists in algebra.
Even if Dirac had known of Clifford's algebras before, this would
not have dimmed the brilliance of the realization that such entities
are of importance for the quantum mechanics of a spinning electron---this
constituting a major and unexpected advance in physical understanding.
\cite[p. 619]{key-10}
\end{quote}
These matrices are appropriate to be introduced in quantum mechanics
since they are like linear operators acting on the wave function similar
to the actions of the non-commuting position and momentum operators.
Rather unexpectedly, these special types of matrices must refer to
and act upon some new degrees of freedom in addition to the usual
position and momentum variables of the quantum particle. The new degrees
of freedom describe the physical spin of the electron and were later
found to be true for all the fermions of nature as well. In the Dirac
equation, these matrices act as operators on the wave function $\psi$
dubbed spinor, a name apparently coined by Paul Ehrenfest, the thesis
advisor of the fortunate duo Uhlenbeck and Goudsmit. What Dirac had
efficaciously introduced in quantum mechanics was a powerful new formalism,
known today as spinor calculus. The very name spinor invokes rotation.
But how exactly the electron spins remains mysterious \cite{key-11}.

Again, using the energy and momentum operators (\ref{eq:2}), the
Dirac equation for a wave function $\psi(x,y,z,t)$ becomes
\begin{widetext}
\begin{equation}
\imath\hbar\frac{\partial}{\partial t}\psi(x,y,z,t)=\left[m_{0}c^{2}\alpha_{0}-\imath\hbar c\left(\frac{\partial}{\partial x}\alpha_{1}+\frac{\partial}{\partial y}\alpha_{2}+\frac{\partial}{\partial z}\alpha_{3}\right)\right]\psi(x,y,z,t)\label{eq:14}
\end{equation}
\end{widetext}
This is the equation essentially in the form originally proposed by
Dirac \cite[p. 255]{key-12}. It is rather helpful for an intuitive
understanding of the Dirac equation. Equation (\ref{eq:14}) can be
presented very elegantly by drawing on the momentum four-vector with
its magnitude remaining invariant under Lorentz transformation. As
usual, we start with the relativistic energy-momentum relation (\ref{eq:1}),
which after rearranging terms becomes
\begin{equation}
\frac{E^{2}}{c^{2}}-p^{2}=m_{0}^{2}c^{2}
\end{equation}
or 
\begin{equation}
\left(\frac{E}{c}\right)^{2}-p_{x}^{2}-p_{y}^{2}-p_{z}^{2}=m_{0}^{2}c^{2}.\label{eq:17}
\end{equation}
Equation (\ref{eq:17}) represents the equation for 4-momentum, whose
magnitude remains invariant under Lorenz transformation. By using
Dirac $\gamma$ matrices,
\begin{alignat}{1}
\gamma_{0} & =\left(\begin{array}{cccc}
1 & 0 & 0 & 0\\
0 & 1 & 0 & 0\\
0 & 0 & -1 & 0\\
0 & 0 & 0 & -1
\end{array}\right),\\
\gamma_{1} & =\left(\begin{array}{cccc}
0 & 0 & 0 & 1\\
0 & 0 & 1 & 0\\
0 & -1 & 0 & 0\\
-1 & 0 & 0 & 0
\end{array}\right),\\
\gamma_{2} & =\left(\begin{array}{cccc}
0 & 0 & 0 & -\imath\\
0 & 0 & \imath & 0\\
0 & \imath & 0 & 0\\
-\imath & 0 & 0 & 0
\end{array}\right),\\
\gamma_{3} & =\left(\begin{array}{cccc}
0 & 0 & 1 & 0\\
0 & 0 & 0 & -1\\
-1 & 0 & 0 & 0\\
0 & 1 & 0 & 0
\end{array}\right).
\end{alignat}
where
\begin{equation}
\gamma_{0}^{2}=1,\quad\gamma_{1}^{2}=\gamma_{2}^{2}=\gamma_{3}^{2}=-1,
\end{equation}
\begin{equation}
\gamma_{i}\gamma_{j}=-\gamma_{j}\gamma_{i}\qquad(i\ne j)
\end{equation}
we construct an equation
\begin{equation}
\gamma_{0}\frac{E}{c}+\gamma_{1}p_{x}+\gamma_{2}p_{y}+\gamma_{3}p_{z}=m_{0}c\label{eq:20}
\end{equation}
where by squaring equation (\ref{eq:20}), we can get equation (\ref{eq:17}).

Again, using the energy and momentum operators (\ref{eq:2}), equation
(\ref{eq:20}) becomes
\begin{equation}
\imath\hbar\left(\gamma_{0}\frac{1}{c}\frac{\partial}{\partial t}-\gamma_{1}\frac{\partial}{\partial x}-\gamma_{2}\frac{\partial}{\partial y}-\gamma_{3}\frac{\partial}{\partial z}\right)=m_{0}c,
\end{equation}
which in vector multiplication notation can be written as
\begin{equation}
\imath\hbar\left(\gamma_{0}\frac{1}{c}\frac{\partial}{\partial t}-\gamma.\nabla\right)=m_{0}c.\label{eq:23}
\end{equation}
The quantity within parenthesis in equation (\ref{eq:23}) is called
the \emph{Dirac operator} that can be written as $\gamma^{\mu}\partial_{\mu}$
where $\gamma^{\mu}$ are Dirac matrices and $\partial_{\mu}$ is
the relativistically invariant 4-gradient,
\begin{equation}
\partial_{\mu}=\left(\frac{1}{c}\frac{\partial}{\partial t},\nabla\right).
\end{equation}
Then equation (\ref{eq:23}) can be written elegantly as
\begin{equation}
\imath\hbar\gamma^{\mu}\partial_{\mu}=m_{0}c
\end{equation}
With the Dirac operator operating on a wave function $\psi$, we get
an eigenvalue equation
\begin{equation}
\imath\hbar\gamma^{\mu}\partial_{\mu}\psi-m_{0}c\psi=0
\end{equation}
Using natural units $\hbar=c=1$ and the \emph{Feynman slash notation}
$\slashed{\partial}=\gamma^{\mu}\partial_{\mu}$, the equation becomes
\begin{equation}
\left(\imath\slashed{\partial}-m_{0}\right)\psi=0
\end{equation}
This is one of the most graceful equations, which is just about as
well known for its stunningly magnificent beauty as for its dramatic
prediction for the deeper secrets of nature. Its physical consequences
are more extensive and far-reaching than anyone could have imagined.
A great deal more was hidden in the Dirac equation than even the author
himself anticipated.

\section{Solution of Dirac Equation}

Since the Hamiltonian is a $4\times4$ matrix, the wave-function $\psi(r,t)$
it acts on is naturally a 4-component column vector: 
\begin{equation}
\psi(r,t)=\left(\begin{array}{c}
\psi_{1}(r,t)\\
\psi_{2}(r,t)\\
\psi_{3}(r,t)\\
\psi_{4}(r,t)
\end{array}\right)
\end{equation}
which gives a set of four coupled linear equations. The four-component
wave function represents a new class of mathematical object in physical
theories and makes their first appearance in Dirac's masterpiece.

Shortly after its earliest presentation \cite{key-13}, Dirac equation
for the electron was solved for the hydrogen atom by Gordon \cite{key-14}
and Darwin \cite{key-15} and subsequently by others. Now the results
appear in contemporary textbooks \cite{key-16,key-17}. A summary
of the results is presented below:

(a) Dirac's initial goal of developing a quantum theory of the electron
with positive definite probability in a relativistic quantum theory
was achieved with a first order time derivative.

(b) As an immediate consequence of his equation, the paradox of the
mysterious quantized spin and the associated magnetic moment of the
electron were understood to be a natural outcome of the Dirac equation.
In the rest frame, the spin operator $\hat{s}$ for the Dirac particle,
having eigenvalues $\pm\frac{1}{2}\hbar$, necessarily has an intrinsic
spin, $s=\frac{1}{2}$, that is not related to ordinary orbital angular
momentum~$\ell$. However, the spin operator does not commute with
the Hamiltonian in any frame other than the rest frame, $p=0$, so
the expectation value of the spin operator is not a conserved quantity
for $p\ne0$. This agrees with the Stern--Gerlach experiment, where
the single electron in the outermost shell having $\ell=0$, $m=0$,
still shows deflection of the beam into only two discrete values of
its magnetic moment relating to $s=\pm\frac{1}{2}$ and confirming
that spin is an intrinsic property of the electron that was enigmatic
even to Pauli himself.

Similarly, the orbital angular momentum operator $\hat{L}$ does not
commute with the Hamiltonian in any frame other than the rest frame,
so orbital momentum is not a conserved quantity either. However, the
operator \mbox{$\hat{J}=\hat{L}+\hat{s}$} commutes with the Hamiltonian
in all frames. This strongly suggests that $\hat{J}$ should be interpreted
as the operator for the total angular momentum, which is conserved.

(c) The gyro magnetic ratio of the electron spin turned out to be
very close to $2$. But its spin angular momentum $\frac{1}{2}\hbar$
compensates giving an effective value of $g=1$ as previously assumed.

(d) The solution of the problem of the central field, which can be
carried out exactly, leads to the prediction of the fine details of
the spectrum of the hydrogen atom in rigorous agreement with the experimental
results.

(e) The two unexpected extra equations resulting from the four
component wave function seemed very perplexing. Two equations were
enough to explain the two values of the spin as in Pauli's earlier
provisional theory.

What then the other two spin values signify? Furthermore, the two
extra equations seem to represent negative energy states. Even classically
the negative energy states are indicated when we take the square root
of equation (\ref{eq:1}). They are usually discarded as non-physical.
However, doing so in quantum mechanics could be risky since complex
numbers are used. But accepting them would make transitions from positive
to negative energy states possible. Seemingly out of desperation,
Dirac proposed his uncanny sea of electrons. Presumably all the possible
states of the electron in the universe would be filled following Pauli's
exclusion principle. One missing electron from the sea would be a
hole into which an electron could transition. In spite of its obvious
criticism, this strange idea persisted for a rather long time until
Julian Schwinger proposed to retire it as an historical oddity.

Dirac then toyed with the idea that the two negative energy states
could be for protons, which were the only positively charged particles
known at the time. Evidently that idea also ran into trouble since
the protons are about two thousand times heavier than the electrons.
Dirac finally hit upon an unexpected winner. Seemingly the very baffling
problem at its inception turned out to be a colossal discovery hidden
in the equations. After the early confusion, a startling corollary
of Dirac's equation yielded the prediction \cite{key-18} of a new
particle, the positron, an antimatter counterpart of the electron!
The spins in the two additional equations belonged to two positrons.

The new particle would have the same mass as the electron but with
opposite electric charge, and be capable of annihilating an electron,
both being converted into pure energy in the process. Conversely,
an electron and positron pair could be created provided there was
enough energy available, unveiling a totally new vista of nature's
mysteries. Exactly such an occurrence was soon discovered by Carl
Anderson in 1932 from a meticulous analysis of cosmic ray tracks in
a cloud chamber \cite{key-19}.

The physics Laureate Steven Weinberg affirms
\begin{quote}
The discovery of the more-or-less predicted positrons, together with
the earlier success of the Dirac equation in accounting for the magnetic
moment of the electron and the fine structure of hydrogen, gave Dirac's
theory a prestige that it has held for over six decades. \cite[p. 13]{key-20}
\end{quote}
And Franck Wilczek articulates
\begin{quotation}
Dirac searched for a mathematical equation satisfying physically motivated
hypotheses. He found that to do so he actually needed a system of
equations, with four components. This was a surprise. Two components
were most welcome, as they clearly represented the two possible directions
of an electron's spin. But the extra doubling at first had no convincing
physical interpretation. Indeed, it undermined the assumed meaning
of the equation. Yet the equation had taken on a life of its own,
transcending the ideas that gave birth to it, and before very long
the two extra components were recognized to portend the spinning positron,
as we saw.

With this convergence, I think, we reach the heart of Dirac's method
in reaching the Dirac equation, which was likewise Maxwell's in reaching
the Maxwell equations, and Einstein's in reaching both the special
and the general theories of relativity. \cite[p. 185]{key-21}
\end{quotation}
Is it then any wonder that Dirac's accomplishment has been described
as fully on a par with the works of Newton, Maxwell, and Einstein
before him \cite[p. 228]{key-22}.

\section{Harbinger of Quantum Field Theory}

While there are many aspects of Dirac's equation that led to uncovering
the riddles of nature, we will now focus on how his seminal work initiated
the development of the Quantum Field Theory that steers us to the
deeper mysteries of nature revealed to date.

The groundbreaking discovery of creation and annihilation of particle--antiparticle
pairs indeed led to a radical change in our perception of nature,
guiding us to the conclusion that particles could not be the primary
reality of the universe as had been intuitively presumed so far. What
then is the primary reality? Again, Dirac's pioneering contribution
eventually guided us to the answer. As Anthony Zee articulates
\begin{quote}
It is in the peculiar confluence of special relativity and quantum
mechanics that a new set of phenomena arises: Particles can be born
and particles can die. It is this matter of birth, life, and death
that requires the development of a new subject in physics, that of
quantum field theory. \cite[p. 4]{key-23}
\end{quote}
The very foundation of our understanding of the deeper secrets of
nature is thus provided by successfully combining relativity with
quantum mechanics in a way that even Dirac did not foresee when he
embarked on the project.

Since the properties of a particle like an electron are exactly the
same throughout the universe, a natural inference was that an underlying
space filling quantum field is the primary reality, a quantized excitation
of which appears as a particle. Thus Dirac's pivotal discovery eventually
led to the new paradigm of Quantum Field Theory (QFT) that presents
a unified view of nature, obliterating our long held perception that
matter particles and forces possess totally different attributes.

Additionally, it was Dirac's innovative work leading to the first
successful quantization of the electromagnetic field \cite{key-24}
that also gave a thoroughly quantum-mechanical treatment of spontaneous
emission. Furthermore, Dirac also coined the name Quantum Electrodynamics
for the simplest quantum field theory. Unfortunately, the theory got
hopelessly bogged down for quite some time due to unexpected difficulties
arising from the lack of cancellation of infinities.

Motivated by some key new experimental observations, particularly
those of Willis Lamb \cite{key-25} and P. Kusch \cite{key-26}, a
new generation of young scientists predominantly Julian Schwinger
\cite{key-27}, Richard Feynman \cite{key-28}, Shinichiro Tomonaga
\cite{key-29} and Freeman Dyson \cite{key-30} succeeded in eliminating
the infinities by utilizing the technique of renormalization. Following
this achievement, it was natural to make attempts to develop a Quantum
Field Theory that would encompass not only photons, electrons and
positrons but also the plethora of other particles that were continually
being discovered at the time in particle accelerators and cosmic rays
together with the weak and strong nuclear forces that act upon them.

With contributions from many outstanding physicists, some of whom
garnered Nobel Prizes for their exceptional efforts throughout the
second half of the twentieth century, we now have such a quantum field
theory, known as the Standard Model of particle physics, which is
a comprehensively expanded version of the quantum electrodynamics
initiated by Paul Dirac. Experimentally verifiable accounts of the
deepest design of nature revealed to date rest upon this quantum field
theory of Standard Model. Needless to say, it by no means provides
a complete account. Most conspicuously, it does not include gravity
nor does it account for dark matter, dark energy, and possibly others
that we have not glimpsed yet. However, Steven Weinberg affirms
\begin{quote}
We can be certain that the Standard Model will appear as at least
an approximate feature of any better future theory. \cite[p. 264]{key-31}
\end{quote}
Based on the discoveries completed to date, we now make a plausible
effort to understand the various enigmas of the quantum world. Quantum
theory is the most successful and supremely accurate theory propounded
in recorded history. Yet, famously, certain aspects of the theory
appear as weird and counter intuitive. Fortunately, I believe this
``weirdness'' can be understood from the precepts of QFT that portrays
a reality different from our daily classical reality, both coexisting
by transitioning from the quantum to the classical domain.

\section{The wave function of an electron}

According to QFT, a particle like an electron arises as a quantized
excitation of the underlying electron quantum field. Such an energy-momentum
eigenstate of the field can be expressed as a specific Lorentz covariant
superposition of field shapes of the electron field along with all
the other quantum fields of the Standard Model of particle physics.
Superposition of field shapes in a one-particle state is not stationary
in time but evolve in a simple wavelike manner.

The individual field shapes, each with their own computable dynamic
time evolution, are actually the vacuum fluctuations comprising the
very structure of the energy-momentum eigenstate. The vacuum fluctuations
are evanescent in the sense that they pass away soon after coming
into being. But new ones are constantly boiling up to establish an
equilibrium distribution so stable that their contribution to the
electron $g$-factor results in a measurement accuracy of one part
in a trillion \cite{key-32}. The Lorentz covariant superposition
of vacuum fluctuations of all the quantum fields in the one-particle
quantum state can therefore be conveniently represented using the
Fourier synthesis approach leading to a well behaved smooth wave packet
that is everywhere continuous and continuously differentiable.

The wave function $\psi(x)$, for simplicity in one dimension, will
be given by the Fourier integral
\begin{equation}
\psi(x)=\frac{1}{\sqrt{2\pi}}\int_{-\infty}^{\infty}\tilde{\psi}\left(k\right)e^{\imath kx}dk
\end{equation}
where $\tilde{\psi}(k)$ is a function that quantifies the amount
of each wave number component $k=\frac{2\pi}{\lambda}$ that gets
added to the combination.

From Fourier analysis, we also know that the spatial wave function
$\psi(x)$ and the wave number function $\tilde{\psi}(k)$ are a Fourier
transform pair. Therefore, we can find the wave number function through
the Fourier transform of $\psi(x)$ as
\begin{equation}
\tilde{\psi}\left(k\right)=\frac{1}{\sqrt{2\pi}}\int_{-\infty}^{\infty}\psi\left(x\right)e^{-\imath kx}dx
\end{equation}
Thus the Fourier transform relationship between $\psi(x)$ and $\tilde{\psi}(k)$,
where $x$ and $k$ are known as conjugate variables, can help us
determine the frequency or the wave number content of any spatial
wave function. In order to determine the time evolution of the wave
packet function, we need to incorporate the time term to the spatial
function. Accordingly,
\begin{equation}
\psi(x,t)=\frac{1}{\sqrt{2\pi}}\int_{-\infty}^{\infty}\tilde{\psi}\left(k\right)e^{i\left(kx-\omega t\right)}dk
\end{equation}
where $\omega=2\pi\nu$ is the angular frequency.

A wave packet or a wave function is localized and therefore can represent
a quantum particle, but just \emph{holistically}, since only the totality
of the wave packet represents all the conserved quantities of the
energy-momentum eigenstate of a particle such as mass, charge, and
spin. Creation of a similar wave function of a particle has been presented
earlier \cite{key-33} from a rather exploratory view for a better
intuitive understanding.

When quantum mechanics with its hallmark wave function supervened
on the atomic domain of physics and commenced explaining its mysterious
workings with uncanny consistency, it appeared totally contrary to
our intuition developed from classical physics. Even a century later,
quantum mechanics still perplexes most people including many scientists.
However, there appear to be plausible answers to the riddles of quantum
mechanics based on the discoveries of the quantum field theory of
the Standard Model, again the crown jewel of physics that was forged
through, commencing with Paul Dirac's uncanny insight.

Following is a consolidated account of resolutions of the perplexities
of quantum mechanics, some of which has been presented in parts by
the author in previous publications \cite{key-34,new-1,new-2}. For
simplicity, here we will be dealing solely with a single quantum particle,
a multi particle extension of which is under preparation and will
be presented later. Since the depiction of the single particle wave
function presented here is based on a fully relativistic account,
its reality should be extendable to multi particle systems. Although
some emergent properties like quantum entanglement appear in a multi
particle system, the fundamental quantum realism of a single particle
wave function should shore up the reality of the multi particle wave
function, which for expediency is presented in a rather abstract Hilbert
space formalism.

\subsection{\label{sub:reality}Reality of the Wave Function}

From the early days of quantum mechanics, physicists have argued about
the reality of the wave function. It is surprising to find them still
debating it, particularly in light of Erwin Schr\"{o}dinger's demonstration
that the mysteriously discrete Bohr orbitals of the hydrogen atom
arise from standing wave patterns of the wave function of the electron.
Even after almost a century, it is not uncommon for very notable contemporary
scientists to pronounce the wave packet for a particle like an electron
to be nothing but a fictitious mathematical construct. Does not it
stretch credibility to imagine that fictitious waves can possibly
make such unmistakably robust standing waves to form discrete orbitals
of atoms?

It is also hard to fathom how anyone can doubt the reality of a wave
associated with a particle after an unambiguous experimental demonstration
of the electron diffraction pattern of the de Broglie matter wave.
Although de Broglie himself failed to recognize the reality of the
observed wave, he correctly surmised the origin of the wave to be
a result of the relativistic effect of an unspecified internal frequency
associated with the particle. Now we know that the internal frequency
relates to the relativistic equation: $\nu=\frac{m_{0}c^{2}}{h}$
indicating that a matter particle is itself a wave packet as is unmistakably
evident from QFT.

Historically, the de Broglie hypothesis that a matter particle behaves
like a wave of length, $\lambda$, related to its momentum, $p$,
through the Planck constant, $h$ to be $\lambda=\frac{h}{p}$ or
equivalently $p=\hbar k$ helped Schr\"{o}dinger to develop his wave
equation that supercharged the brand new discipline of quantum mechanics.
Using the Lorentz invariant property of the wave packet, the relation
$p=\hbar k$ can indeed be derived irrespective of the mass of the
particle \cite{key-34} and therefore need not be a mere hypothesis.

As described earlier, a quantum particle is a holistic wave packet
consisting of irregular disturbances of the various quantum fields
of the Standard Model and these disturbances are not readily measurable.
Their distinct consequences, however, are by now well established
for explaining the Lamb shift and the anomalous magnetic moment of
the electron. These disturbances of the quantum fields otherwise known
as vacuum fluctuations are like the `beables' eponymously coined by
John Stuart Bell that corresponds to something that really exists
physically but not directly observable. In Bell's words
\begin{quote}
The beables of the theory are those elements which might correspond
to elements of reality, to things which exist. Their existence does
not depend on `observation'. \cite[p. 174]{key-35}
\end{quote}
Quarks and gluons are distinct examples of beables. No one can see
an isolated quark or a gluon, but few scientists doubt their real
existence.

Some graphic depiction of the vacuum fluctuations could as well be
offered soon. The 2014 Kavli Prize in Astrophysics was awarded for
pioneering contributions to the theory of cosmic inflation. The predictions
of the simplest versions of the theory have been so successful that
most cosmologists accept that some form of inflation truly did occur
in the very early universe. Assuming the veracity of cosmic inflation
at the dawn of the universe, the imprint of the disturbances of the
quantum field could manifestly be discernible as immensely enlarged
vacuum fluctuations in the cosmic microwave background radiation anisotropy,
recorded both by the WMAP and Planck satellite \cite{key-36}.

Since the quantum fields manifestly comprise a reality of the universe,
their disturbances should be no less real. Thus the mere straightforward
immeasurability of the disturbances of the quantum fields does not
make them fictitious, providing distinct validation for the wave function
portrayed here to be objectively real. There is every reason to believe
that the very weave of our universe supports the objective reality
of the depicted wave function, which represents a natural phenomenon
and not just a mathematical postulate.

The perplexing wave-particle duality should by now be comprehensible.
In reality, there is no such thing in the quantum domain as a classical
particle akin to a tiny marble ball. What we call a quantum particle
is actually a wave packet consisting of real field disturbances that
in their totality behave as a particle. The idea of a particle, however,
is so ingrained in our awareness that it is unlikely to fade away
anytime soon. This is especially true since at energies higher than
that corresponding to the Compton wavelength, the exchange of momentum
is more particle-like, as in inelastic scattering. Also the visible
tracks of the high energy elementary particles in a particle detector
like a cloud chamber are particle-like.

\subsection{The Uncertainty Principle}

On the basis of those characteristics of the wave function that represent
a quantum particle, the mysterious uncertainty principle of quantum
physics can now be understood as a perfectly natural incidence. The
Fourier transform correlations between conjugate variable pairs of
any wave packet have powerful consequences since these variables obey
the uncertainty relation
\begin{equation}
\Delta x\Delta k\geq\frac{1}{2}
\end{equation}
where$\Delta x$ and $\Delta k$ relate to the standard deviations
$\sigma_{x}$ and $\sigma_{k}$ of the wave packet. This is a completely
general property of a wave packet with a reality of its own and is
in fact inherent in the properties of all wave-like systems. It becomes
important in quantum mechanics because of the real wave nature of
particles having the relationship $p=\hbar k$, where $p$ is the
momentum of the particle. Substituting this in the general uncertainty
relationship of a wave packet, the intrinsic uncertainty relation
in quantum mechanics becomes
\begin{equation}
\Delta x\Delta p\geq\frac{1}{2}\hbar
\end{equation}
This uncertainty relationship has been conflated with a somewhat analogous
observer effect, which advocates that measurement of certain systems
cannot be made without affecting the system. In fact, Heisenberg offered
such an observer effect in the quantum domain as a ``physical explanation''
of quantum uncertainty, an explanation that has since gone by the
name of Heisenberg's uncertainty principle. What the uncertainty principle
actually states, however, is a fundamental property of quantum systems,
and is not a statement about observational indeterminacy as was suggested
by Heisenberg, indeed originally branding it the indeterminacy principle.
In fact, some recent studies \cite{key-37,key-38,key-39} highlight
important fundamental differences between uncertainties in quantum
systems and the limitation of measurement in quantum mechanics.

Einstein's fundamental objection to the Copenhagen view of the wave
packet was its assertion that any underlying reality of the uncertainties
was irrelevant and should be accepted under the veil of complementarity.
We have established that this uncertainty is indeed an essential reality,
governed by wave behavior, traceable in its origin back to the wave-particle
duality first envisioned and steadfastly maintained as a reality by
Einstein, alone among his peers, for over a decade.

\subsection{Born's Rule of Probability}

Perhaps the most perplexing aspect of quantum physics is its probabilistic
prediction, as formalized by Born's rule. From experiments on the
scattering of electrons, Max Born showed that the original suggestion
of Schr\"{o}dinger that the wave packet be considered as the charge distribution
of the electron could not be justified. Instead, Born followed Einstein
in this regard as he stated in his Nobel lecture
\begin{quote}
Again an idea of Einstein's gave me the lead. He had tried to make
the duality of particles (light quanta or photons) and waves comprehensible
by interpreting the square of the optical wave amplitudes as probability
density for the occurrence of photons. This concept could at once
be carried over to the $\psi$-function: $\left|\psi\right|^{2}$
ought to represent the probability density for electrons (or other
particles). \cite{key-40}
\end{quote}
This inference is in accord with the correct nature of the wave function,
as presented in Section \ref{sub:reality}. As the amplitudes of the
wave packet consist of disturbances of various diverse quantum fields,
they could not simply represent a distribution of charge or any other classical
property of the particle. Since the absolute square of the
amplitude is non-negative, whereas the amplitude generally contains
complex numbers, $\left|\psi\right|^{2}$ should represent the probability
density of a quantum particle like an electron. This is why the amplitude
of the wave function is commonly called the \emph{quantum probability amplitude}.
Since the total probability must be one, we have to normalize the
wave function as
\begin{equation}
\int_{-\infty}^{\infty}\left|\psi\left(x,t\right)\right|^{2}dx=1
\end{equation}
Born's rule is inseparably connected to the perplexing measurement
problem. Although its exact mechanism is intensely debated, the most
popular paradigm appears to be that of decoherence, first posited
by H. Deter Zeh \cite{key-41} and extensively studied by Wojciech
Zurek \cite{key-42} and others. In this model the quantum system
to be measured gets entangled with the quantum constituents of the
required macroscopic detector ultimately resulting in the selection
of the measurable event.

Zurek contends \cite{key-43} that the Born rule can actually be derived
from the theory of decoherence as opposed to being a mere postulate
of quantum theory. There is indeed some support for his contention
\cite{key-44} although why a particular probability out of many others
materializes needs to be explained \cite{key-45}. Quite possibly
this is not definitively answerable because of the inherent, irreducible
indeterminacy of the primordial quantum fluctuations that is irrefutably
carried over to the elementary particles.

The baffling `collapse of the wave function', however, has a genuine
explanation based on the nature of the wave function described in
Section \ref{sub:reality}. The holistic feature of the wave function
requires that the entirety of the wave function be acquired all together
or not at all. This is aided by the fact that the wave function of
the particle is also entangled with space \cite{key-46}. Hence, the
spread-out wave function can instantaneously converge at the particular
location of detection.

An example from cosmic history is worth examining in this regard.
The universe about 380,000 years after the big bang consisted primarily
of hydrogen ions (protons) and electrons, along with neutral helium
atoms. An electron would naturally be attracted to the proton, starting
to emit electromagnetic radiation due to its motion. But a much more
rapid process would take place when the electron, while aligned in
the direction of the proton, spontaneously emits a photon with an
amount of energy that exactly matches the potential energy of the
electron and an orbital of the hydrogen atom. In this process the
wave function of the electron can directly wind up as the wave function
of the particular orbital of the hydrogen atom without having to undergo
a typical collapse to any particular point. Such episodes would reveal
that the wave function does not necessarily always need to go through
a traditional collapse for detection.

The above episode, which indeed took place in a universal scale, could be used as a pragmatic toy model for grasping the much discussed but yet to be resolved measurement problem. In this case, the necessary macroscopic detection device consists solely of essentially an assembly of the same quantum constituents, namely, atomic hydrogen ions. Also, the quantum state to be detected is a simple position wave function of the electron. Even without going into all the details, one can readily comprehend that in addition to predicting only the probability of detection instead of a certainty, there is also an inherent, irreducible element of indeterminacy in the occurrence of any particular probability since the predominant necessary event of the spontaneous emission of a photon of matching potential energy by the electron is totally unpredictable from the Born rule.

The counter-intuitive simultaneous presence of the electron in more
than one place can also be understood in terms of the nature of the
wave function. Since the wave function is spread out in space, which
can be substantial due to dispersion, the probability of finding the
particle in multiple places will be significant. Again, in each location
the wave function has to be taken in its entirety, only that represents
all the properties of the particle, thus giving an ostensible appearance
of the electron in many places at the same time.

As a manifest evidence of the wave-particle duality, the
famous double slit experiment, in which one quantum particle at
a time is sent through two adjacent slits and results in a wave-like
interference pattern, also has a natural explanation. The wave function
consisting of linear superposition of amplitudes gets divided into
two parts at the slits, which then interferes resulting in a typical
wave diffraction pattern, yet detected only at a point with the characteristic
totality of the wave packet. The amplitude of the wave packet can
be split, but the square of the amplitude of the entire wave packet
required for detection cannot be divided. Thus, each individual electron
contributes one dot to an overall pattern that looks like the two-slit
interference pattern of a wave. A distinctly typical interference
pattern arises when a sufficient number of quantum particles are sent
through the two slits either one at a time or all of them together
at the same time.

\section{Conclusions}

We have expounded how Dirac's revolutionary equation
has been invaluable in unveiling some of the bewildering mysteries
of the atomic domain. A totally unexpected outcome of Dirac's equation
of the electron led to the development of the quantum field theory
of the Standard Model of particle physics, leading in turn to the
deepest revelation to date, of nature's most profound secrets. We
have shown that the ensuing quantum field theory originating from
Dirac's pioneering contributions can assist in demystifying many of
the lingering mysteries of quantum physics for both the professionals and the general public.

An objectively real wave function can be garnered from the edicts
of the quantum field theory revealing that there are no solid elementary
particles but only holistic wave packets acting like particles. Amazingly,
the paradigm of the wave function presented here proffers plausible
explanation for the long standing inherent perplexities of the wave-particle
duality, uncertainty principle, de Broglie wave hypothesis, the bizarre
wave function collapse, detecting probabilities instead of certainty.

These quantum oddities arise from our natural instinct to understand
the quantum domain in terms of our daily classical reality. The quantum
field theory reveals that the eventual quantum reality is substantially
different from our familiar classical reality.

The quantum fields fill all space of the entire universe for all times.
The fields have infinite degrees of freedom of creating and annihilating
quantum particles locally at each spacetime element and these degrees
of freedom are indeed always active obeying the energy-time uncertainty
relation even in the complete vacuum. The quantum activities have
the inherent special attribute that each event is totally spontaneous
and completely unpredictable as to when it is going to take place.
This is only a slow motion description of the events. In reality,
there are infinite number of these evanescent events occurring locally
at every spacetime element of the universe, the members of the ensemble
having infinitely different amplitudes and speed.

This unique quantum activity, a feature of each one of the quantum
fields of the Standard Model, being present everywhere, ubiquitously
interact with everything in the quantum domain including elementary
particles, atoms and molecules. The unique features such as the characteristic
quantum superposition as well as some inherent unpredictability could
be an inevitable result. Thus the reality of the quantum domain is
substantially different.

However, the quantum domain transitions into our familiar daily macroscopic
classical domain where nature deals with an innumerable number of
particles that mask the quantum activity with phenomena like decoherence.
Experiments confirm that the characteristic activities of the quantum
domain are revealed when a fairly macroscopic object containing a
trillion atoms is shielded from environmental decoherence as completely
as possible. Thus the quantum reality is markedly different, but coexists
by transitioning into classical reality. Much of the quantum enigmas
disappear when we accept this fact of nature.

\section*{Acknowledgment}
The author wishes to thank Per Kraus, Zvi Bern, Danko~Georgiev, Eric D'Hoker and Joseph Rudnick
for illuminating discussions.

\end{document}